\documentclass[aps,twocolumn,showpacs,floatfix]{revtex4}
\usepackage{amssymb}
\usepackage{graphicx}
\usepackage{amsmath}

\input{tcilatex}
\input{epsf}
\begin{document}

\title{Electronic Structure Calculations of Magnetic Exchange Interactions
in Europium Monochalcogenides}
\author{Xiangang Wan$^{1}$, Jinming Dong$^{1}$, Sergej Y. Savrasov$^{2}$}
\affiliation{$^{1}$National Laboratory of Solid State Microstructures and Department of
Physics, Nanjing University, Nanjing 210093, China\\
$^{2}$Department of Physics, University of California, Davis, One Shields
Avenue, Davis, Ca 95616}
\date{\today }

\begin{abstract}
Using a combination of local spin density and Hubbard 1 approximations we
study the mechansim of exchange interacion in EuX (X=O, S, Se and Te). We
reproduce known experimental results about bulk modulus, critical pressure
for structural phase transition, magnetic ordering temperature, spin--wave
dispersions as well as momentum-- and tempearuture--dependent band shift.
Our numerical results show pressure induced competition between the
hybirization enhanced exchange interaction and Kondo--like coupling in EuO.
Possible ways to enhance T$_{c}$ are discussed.
\end{abstract}

\date{\today}
\maketitle

Europium monochalcogenides EuX (X=O, S, Se and Te) have been studied
extensively since 1961\cite{FM semiconductor}. As the only known examples of
Heisenberg ferromagnetism in nature, EuO and EuS have their Curie
temperatures (T$_{c}$) of 69.15 K and 16.57 K, respectively. On the other
hand, EuSe has a complex magnetic structure at low temperatures, and EuTe is
antiferromagnetic\cite{Review}. It was found that doping of EuO by electrons
results in 100\% spin polarization of the conduction electrons\cite{Exchange
splitting in EuO}, and the material has a colossal magnetoresistance effect
stronger than famous manganites. Moreover, very recently EuO has been
integrated with Si and GaN, making it very attractive for spintronic
applications\cite{Doping EuO}, and the interest to these systems has been
renewed \cite{impurity level in Doped EuO,Band structure,O-2p important 2009
PRL, Pressure EuX, EuX band structure,XMLD EuO,Nano EuS,Gd-doped EuO}.

As finding ways to raise the T$_{c}$ up to room temperatures in EuX is of
both fundamental and technological importance, many past studies of their
magnetic exchange mechanism appeared in the literature. Based on a model
calculation, Kasuya \cite{Kasuya} proposed that the nearest neighbor
exchange coupling $J_{1}$ is induced by the indirect exchange between 4%
\textit{f} and 5\textit{d} electrons of Eu while the superexchange between
the 4\textit{f} states of Eu and \textit{p} electrons of anion can be
ignored. Liu and Lee\cite{Liu and Li} claimed based on their band structure
calculation that anion valence band has an important contribution to both $%
J_{1}$ and $J_{2}$. On the other hand, based on density functional
calculation and Wannier function analysis, Kune\~{s} et.al\cite{Pickett}
emphasized the importance of hybridization between the 4\textit{f} of Eu and
2\textit{p} of O and the associated superexchange interaction. There have
also been other theoretical models \cite{Chomsky} to describe magnetic
exchange mechanism in EuX including the s--f model\cite{s-f model}.
Experimentally, optical spectroscopy finds a considerable 4\textit{f}-5%
\textit{d} mixing and suggests the importance of \textit{f}-\textit{d}
exchange\cite{optical}; the M\"{o}ssbauer experiment emphasizes the effect
of 6\textit{s} band of Eu and supports the \textit{s}-\textit{f} model\cite%
{Mossbauer}; the neutron diffraction stresses the contribution from the
anion \textit{p} shells\cite{Pressure Tc EuS EuSe}; the angle resolved
photoemission (ARPES)\ experiment observes a momentum-dependent
temperature-induced band shift\cite{O-2p important 2009 PRL} and
contributions from Eu 4\textit{f}--5\textit{d} exchange to $J_{1}$ and from
Eu 4\textit{f}--O\ 2\textit{p} exchange to $J_{2}$ while the x--ray
absorption spectroscopy\cite{Pressure EuX} claims that the exchange
interaction is due to \textit{f}--\textit{d} mixing without involvement of
the anion \textit{p} states. It was also found that pressure\cite{Pressure
Tc EuS EuSe, P data,Pressure EuX,Phase phase tran EuO, P EuO 2, Pressure for
EuO, RKKY for Press EuO}, epitaxial strain\cite{EuO epitaxial strain,Strain
EuS} and carrier doping\cite{Doping EuO,Doped EuO 2006 exp} can vary the T$%
_{c}$\ of EuX significantly although still did not reach the values
comparable with the room temperature.

In this work we address the controversial issue of understanding magnetic
exchange mechanism in europium monochalcogenites using a recently developed
linear response approach\cite{PRL 2006} which is based on a combination of
density functional and dynamical mean field theories \cite{DMFTreview}. We
reproduce major experimental results regarding their spin wave dispersions 
\cite{Spin wave EuO EuS}, pressure dependent transition temperatures\cite%
{Pressure Tc EuS EuSe, Pressure for EuO}, and temperature dependent band
structures \cite{O-2p important 2009 PRL}. We provide conclusive theoretical
insights to various contributions to magnetic exchange interactions.

Our electronic structure calculations with the full potential
linearized-muffin-tin-orbital (LMTO) method\cite{FP-LMTO} are done using the
local \textit{spin} density approximation (LSDA) for the conduction
electrons and atomic Hubbard 1 (Hub1) self--energy to approximate localized\
nature of the Eu f--electrons\cite{DMFTreview} with the on--site Coulomb
interaction parameters $U=$7 eV and $J=$1.2 eV\cite{Pressure EuX,Band
structure}. We also check that our results are robust within the reasonable
range of $U$'s\textit{\ }from 6$\ $to 9 eV. With the electronic structure
information, one can evaluate the magnetic interaction $J$ in the Heisenberg
Hamiltonian $H=-\sum_{ij}J_{ij}S_{i}\cdot S_{j}$, based on a magnetic force
theorem\cite{force theorem} that evaluates linear response due to rotations
of magnetic moments\cite{PRL 2006}. This technique has been used
successfully for evaluating magnetic interactions in a series of Mott
insulating oxides\cite{PRL 2006}, cuprate\cite{HTC} and pnictide\cite%
{Pnictides} superconductors.

The ground state properties predicted by our LSDA+Hub1 calculation including
magnetic moments and energy gaps are found to be in agreement with
experiment. Moreover, the obtained exchange splitting of\ conduction band is
about 0.65 eV, which is close to the experimental value 0.60 eV \cite%
{Exchange splitting in EuO}. Since at ambient pressure EuX compounds
crystallize in rock--salt structure, but change to CsCl--type structure at
high pressures\cite{P data},\cite{Phase phase tran EuO}, we perform our
calculations for both NaCl and CsCl type structures for a number of
different volumes. The total energy vs volume, $E(V)$, curves were fitted by
the Murnaghan equation of state (EOS), and the obtained bulk modulus ($B_{0}$%
) together with its first derivative ($B_{0}^{\prime }$) are listed in Table
I. The crystal phase stability is analyzed by evaluating the enthalpy ($%
H=E+PV$), and the phase--transition pressure $P_{c}$ is evaluated from
crossing the $H(P)$ curves. It is found that LSDA alone significantly
overestimates the values of $B_{0}$ and $P_{c}.$ After inclusion of the
correlation effects, our LSDA+Hub1 calculation reproduces the experimental
values successfully. This is shown in Table I. The parameters of EOS and $%
P_{c}$ are not sensitive to $U$'s in the range 6$-$9 eV.

\begin{table}[tbp]
\caption{Comparison between calculated using LSDA+Hub1 method and
experimental values for the bulk modulus $B_{0}$, its first derivative $%
B_{0}^{\prime }$ and phase transition pressure $P_{c}$ in Europium
monochalcogenides. The experimental values are given in parentheses.}%
\begin{tabular}{ccccc}
\hline
& EuO & EuS & EuSe & EuTe \\ \hline
B$_{0}$(GPa) & 105 (118$^{a}$) & 61 (61$^{b}$) & 53 (52$^{b}$) & 43 (40$^{b}$%
) \\ 
B$^{\prime }$ & 3.2 (2.2$^{a}$) & 2.8 & 2.8 & 2.8 \\ 
P$_{c}$(GPa) & 48 (47$^{a}$) & 26 (22$^{b}$) & 17 (15$^{b}$) & 14 (11$^{b}$)
\\ \hline
\end{tabular}
\newline
$^{a}$Ref. \cite{Phase phase tran EuO}; $^{b}$Ref. \cite{P data}.
\end{table}

Using our LSDA+Hub1 method and the magnetic force theorem, we subsequently
evaluate the exchange constants as the integral over the \textit{q} space
using (8,8,8) grid. The magnetic interactions in these materials are usually
characterized by two exchange constants, $J_{1}$ and $J_{2}$, although there
is an argument about longer range interactions\cite{Long range J}. Our
linear response approach allows us to evaluate $J$'s in real space, and the
numerical results confirm that they are short range with the magnetic
coupling further than the second nearest neighbor to be almost equal to
zero. As shown in Table II, our $J_{1}$ and $J_{2}$ are smaller than the
values extracted from neutron scattering experiment \cite{Spin wave EuO EuS}%
, while quite close to the thermodynamic data\cite{Thermal}. Consistent with
the experiment, we obtain that moving from EuO to EuTe, the strength of $%
J_{1}$ decreases, meanwhile $J_{2}$ changes from FM to AFM--like. For EuSe,
the magnitude of $J_{1}$ is smaller than $J_{2}$, but notice that for the
rock--salt structure the number of nearest neighbors and of second nearest
neighbors is 12 and 6, respectively, so there is a competition resulting in
a complex magnetic structure at low temperatures\cite{Review}. For EuTe,
both $J_{1}$ and $J_{2}$ become AFM--like and the ground state changes to AFM%
\cite{Review}. Using the mean--field approximation (MFA), we also estimate
the magnetic ordering temperature. The agreement between the experimental
and calculated $T_{c}$ is good, with the expected overestimate of the
mean--field value. 
\begin{table}[tbp]
\caption{Calculated and experimental nearest neighbor, $J_{1}$, and second
nearest neighbor, $J_{2}$, exchange couplings as well as magnetic transition
temperature for EuX (X=O, S, Se and Te) in units of K. The positive/negative
signs denote ferro/antiferro magnetic coupling and Curie/Neel temperature.}%
\begin{tabular}{cccccc}
\hline
&  & EuO & EuS & EuSe & EuTe \\ \hline
Our results & $J_{1}$ & 0.60 & 0.12 & 0.10 & -0.03 \\ 
& $J_{2}$ & 0.03 & -0.10 & -0.18 & -0.24 \\ 
Thermodynamic \cite{Thermal} & $J_{1}$ & 0.67 & 0.19 & 0.13 & 0.02 \\ 
& $J_{2}$ & -0.06 & -0.08 & -0.12 & -0.16 \\ 
Neutron Scattering\cite{Spin wave EuO EuS} & $J_{1}$ & 0.61 & 0.24 &  &  \\ 
& $J_{2}$ & 0.12 & -0.12 &  &  \\ 
Our results & $T_{c}$ & 81.1 & 19.6 & -5.9 & -19.8 \\ 
Experimental data \cite{Review} & $T_{c}$ & 69.3 & 16.6 & -7.1 & -12.0 \\ 
\hline
\end{tabular}%
\end{table}

Based on the obtained exchange interactions we evaluate the spin-wave
dispersions of EuO along major high symmetry directions. This is shown in
Fig.1 where for comparison we also plot by symbols the results of neutron
scattering measurements\cite{Spin wave EuO EuS}. Our numerical data are
found in good agreement with the experiment which is done on polycrystalline
samples thus accessing direction averaged spin--wave excitations.

\begin{figure}[tbp]
\centering \includegraphics [width=2.5in] {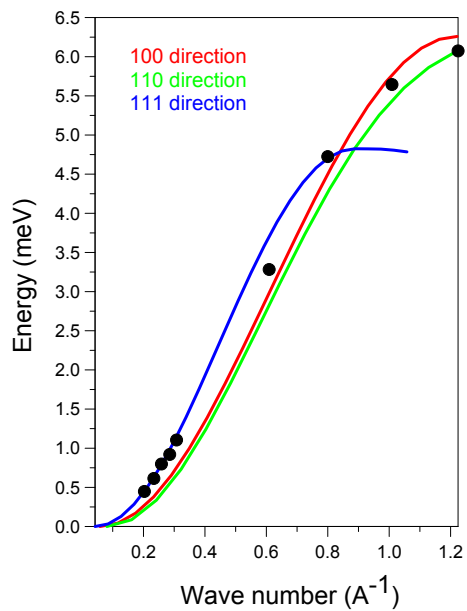}
\caption{Calculated spin--wave dispersions of EuO in comparison with the
experiment (circles)\protect\cite{Spin wave EuO EuS}.}
\end{figure}

One of the controversy about the mechanism of exchange interaction is the
superexchange via the \textit{p} orbital of anion which is believed to be
negligible due to its small hybridization with localized 4\textit{f} orbital%
\cite{Kasuya,Liu and Li}.\ However this viewpoint has been argued due to
recent theoretical works\cite{Pickett,EuO epitaxial strain} and a very
recent observation of the momentum--dependent shift of 4\textit{f} states of
Eu and 2\textit{p} states of O\cite{O-2p important 2009 PRL}. It has even
been suggested that the \textit{f}--\textit{p }superexchange is the leading
factor which induces change of sign in $J_{2}$ when moving from EuO to EuTe%
\cite{EuO epitaxial strain}. To clarify this effect, we performed the
calculation with turned off spin polarization of the conduction band by
using LDA instead of L\textit{S}DA approach. It is interesting that such
LDA+Hub1\ calculation still gives a considerable spin splitting for the 
\textit{p} band of anion, and the magnitude of this splitting is almost the
same as we obtain with the L\textit{S}DA+Hub1 method. Within the LDA+Hub1
framework this spin splitting of the \textit{p} band can only come from the
hybridization with the 4\textit{f} band of Eu. Thus, we confirm that the 
\textit{f}--\textit{p} overlap is not negligible and the spin polarization
in 4\textit{f}\ will result in shifting \textit{p} band as suggested by the
recent experimental work\cite{O-2p important 2009 PRL}.

To further understand the effect of the \textit{f--p} hybridization and in
order to make comparisons with recent ARPES data \cite{O-2p important 2009
PRL} regarding the temperature and moment--dependent Eu-4\textit{f} and O-2%
\textit{p} band shifts we perform the LSDA+Hub1 calculation at temperatures
below and above $T_{c}$. Agree with the experiment\cite{O-2p important 2009
PRL}, our calculation shows that at low temperature (5 K), O-2\textit{p}
state have considerable spin splitting as shown in Fig.2(a). With increasing
temperature the spin splitting decreases and eventually becomes zero when
temperature becomes above T$_{c}$ as shown in Fig2.(b). We also reproduce
the temperature and moment--dependent Eu-4\textit{f} band shift. As shown in
Fig.2, our calculation predict that from low temperature to high
temperature, the top of the Eu-4\textit{f} band at $\Gamma $\ point and 
\textit{X} point have the shift of 0.35 and 0.18 eV, respectively. These
values are in good agreement with the experimental values of 0.32 and 0.07 eV%
\cite{O-2p important 2009 PRL}.

\begin{figure}[tbp]
\centering \includegraphics [height=2.5 in] {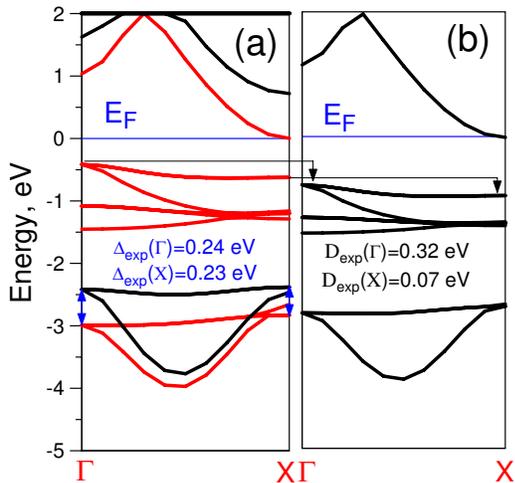}
\caption{Temperature dependent LSDA+Hub1 band structure of EuO. (a)
Temperature is 5 K. The red and black lines denote the majority and minority
spin state respectively. (b) temperature is above T$_{c}$. By arrows we show
the experimental spin splitting in the 2\textit{p} O band and the momentum
dependent shift for the top of the 4\textit{f} Eu band.}
\end{figure}

Based on these arguments and noticing that EuX has the NaCl--type structure
with anion\ located between two Eu atoms, one can then expect that the 4%
\textit{f--}2\textit{p}--4\textit{f} superexchange should play an important
role in the exchange mechanism. However, our values of $J_{1}$ and $J_{2}$
extracted from LDA+Hub1 calculation are coming out to be very small (e.g.,
for EuO, both $J_{1}$ and $J_{2}$ $\lesssim $ 0.001 K) as compared to L$S$%
DA+Hub1 result shown in Table II. This emphasizes that the major part of the
exchange process is going thru virtual excitations to the conduction band.
If one does the same test for Mott insulating oxides\cite{PRL 2006} and
cuprates\cite{HTC} where the 3\textit{d}-2\textit{p--}3\textit{d }%
superexchange dominates, one finds that the results of the calculations with
LDA+Hub1 and L\textit{S}DA+Hub1 are indeed similar to each other. This
clearly indicates that the spin polarization in the conduction band is
essential to explain the magnetic behavior of EuX. At the same time one can
also conclude that there is considerable \textit{f}--\textit{p}
hybridization in EuX system as suggested by recent experiment\cite{O-2p
important 2009 PRL} and theory work\cite{Pickett,EuO epitaxial strain}.
However the \textit{f}--\textit{p} superexchange being the second order
effect can be ignored for those systems.

To see which exactly orbitals are contributing to the exchange process, we
additionally made a calculation of $J$'s with an artificial external
potential applied to a particular orbital. It turns out that a downshift of 5%
\textit{p} orbital for Eu or of \textit{s} orbitals for anion does not
affect the exchange constants. On the other hand, they are extremely
sensitive to the position of 5\textit{d} states of Eu: even a small
up--shift here significantly decreases the strength of $J_{1}$ and $J_{2}$
due to the increase of band separation between 4\textit{f} and 5\textit{d}
and following 5\textit{d--}4\textit{f} dehybridization. Shifting 6\textit{s}
level of Eu affects the $J_{1}$ and $J_{2}$ as well, although the effect is
smaller than for 5\textit{d}. Since anion atoms are between second nearest
neighbor Eu ions, it is natural to expect that $J_{2}$ is mediated by the 2%
\textit{p} electrons of anion\cite{O-2p important 2009 PRL,Pressure Tc EuS
EuSe} through the \textit{d--p} and \textit{s}--\textit{p} hybridization. As
a result the value of $J_{2}$ should be sensitive to the position of the 2%
\textit{p} band. Our numerical calculation, on the other hand, shows that
both $J_{1}$ and $J_{2}$ are almost insensitive to the shift of the \textit{p%
} band of anion. This result is consistent with our previous conclusion on
the smallness of 4\textit{f}-2\textit{p}-4\textit{f} superexchange. For
example, we upshift the \textit{p} level of O in EuO by 1.0 eV but $J_{1}$
and $J_{2}$ are almost unchanged. So, our calculation clearly rules out
possible contributions from the \textit{d}--\textit{p} and \textit{s}--%
\textit{p} hybridization, and even the second nearest neighbor coupling $%
J_{2}$ is mediated by the 6\textit{s} and 5\textit{d} electrons.

\begin{figure}[tbp]
\centering \includegraphics [height=2.5in] {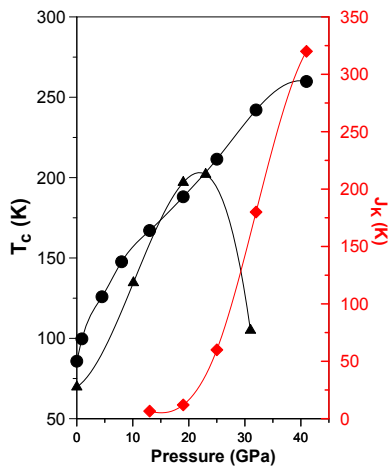}
\caption{Pressure dependence of magnetic transition temperature of EuO
experimental (triangle)\protect\cite{Pressure for EuO}, theoretical(circles)
as well as Pressure dependence of Kondo coupling J$_{K}$.}
\end{figure}

We have studied the effect of pressure which does not only enhance the
hybridization between the conduction band and the $f$ states, but also
increases the crystal--field splitting between 5$d$ $t_{2g}$ and $e_{g}$
states of Eu and reduces the energy gap between the $t_{2g}$ and the 4$f$%
\textit{\ }states\cite{Pressure Tc EuS EuSe}. Consistent with the
experiment, our theoretical results show that pressure enhances the exchange
constants and $T_{c}$ of EuS, while it changes the ground state of EuSe and
EuTe from AFM to FM. However, for EuO our numerical data agree with the
experiment only at the low pressure region, while for pressures larger than
20 GPa, our calculated Curie temperature still rises but the experimental
one decreases\cite{Pressure for EuO}. This is illustrated in Fig.3. Our
calculation shows that pressure closes the band gap and results in metallic
behavior of EuO, but even for pressures as high as 40 GPa, the calculated
exchange coupling is still short range. Based on our LSDA+Hub1 calculation,
we, on the other hand, estimate the on--site Kondo coupling strength $J_{K}$
using a method described by us earlier\cite{Jk}. As shown in Fig.3, $J_{K}$\
increases rapidly above 20 GPa, which results in not only the Kondo--like
screening but also in an AFM\ like intersite coupling in the second order
perturbation theory with respect to $J_{K}$. Both effects will suppress the $%
T_{c}$ as is seen experimentally\cite{Pressure for EuO}.

We have finally studied the effect of electron doping on $T_{c}$ for EuO
using the virtual crystal approximation (VCA). Our calculation shows that
there are two competing factors: since the bottom of the conduction band
consists mainly of the majority spin, the doped electron will enter the
spin--polarized manifold, and this results in the onset of moment in 5%
\textit{d} band\ which will enhance $T_{c}$. On the other hand, free
carriers will induce the RKKY interaction which will supress $T_{c}$. We
indeed find that our theoretical $T_{c}$ first increases as the function of
doping , and then goes through a maximum in accord with the experimental
trend\cite{Doped EuO 2006 exp}. So we conclude that a combination of both
doping and pressure can be the efficient ways to reach higher $T_{c}\ $in
those systems.

In summary, we have calculated the exchange constants of EuX and reveal
contributions from various orbitals relevant to the exchange mechanism. In
particular, we showed that the 5\textit{d--}4\textit{f} and 6\textit{s}--4%
\textit{f} indirect exchange is dominant while the \textit{p} electrons of
anion has no contribution to it. We also reproduced that bulk modulus,
pressure--induced phase transition and magnetic ordering temperatures as
well as the spin--wave dispersions agree well with the experiment. We also
suggest that the abnormal behavior of pressure dependent $T_{c}$ in EuO is
due to increase in Kondo--like coupling in the metallic phase.

The work was supported by National Key Project for Basic Research of China
(Grant No. 2006CB921802, and 2010CB923404), NSFC under Grant No. 10774067,
and 10974082. We also acknowledge the DOE NEUP subcontract \#88708 and KITP
where part of the work has been performed.

\end{document}